\documentclass[twoside,twocolumn,floatfix,pra]{revtex4}

\usepackage{float}
\usepackage{graphicx}
\usepackage{pstricks}
\usepackage{pst-grad}
\usepackage{pst-node}
\usepackage{multido}
\usepackage{pst-plot}
\usepackage{psfrag}
\usepackage[latin1]{inputenc}
\usepackage{amssymb}

\newcommand{\bra}[1]{\ensuremath{\langle#1|}}
\newcommand{\ket}[1]{\ensuremath{|#1\rangle}}

\newcommand{\up}{\!\! \uparrow}
\newcommand{\down}{\!\!\downarrow}
\newcommand{\eff}{\textrm{eff}}
\newcommand{\ch}{\mathcal C}
\newcommand{\pout}{\textrm{out}}
\newcommand{\pin}{\textrm{in}}
\newcommand{\pos}{ \hat L }

\newcommand{\groc}[1]{\pscircle[linewidth=1pt, fillstyle=gradient, gradmidpoint=0, gradangle=45, gradbegin=yellow, gradend=white, shadow=true, shadowcolor=gray, shadowsize=1.5pt](#1){.45}}
\newcommand{\qubit}[1]{\pscircle[linewidth=1pt, fillstyle=gradient, gradmidpoint=0, gradangle=45, gradbegin=blue, gradend=lightgray, shadow=true, shadowcolor=gray, shadowsize=1.5pt](#1){.45}}
\newcommand{\none}[1]{\pscircle[linewidth=1pt, linestyle=dashed, dash=1pt 1pt, fillstyle=gradient, gradmidpoint=0, gradangle=45, gradbegin=white, gradend=lightgray](#1){.45}}

\newcommand{\eref}[1]{(\ref{#1})}

\begin{document}







\title{Quantum Information Processing with Quantum Zeno Many-Body Dynamics}
\author{Alex Monras}
\affiliation{Dipartemento di Matematica e Informatica, Universit\`a degli Studi di Salerno, Fisciano (SA), Italy\\
The School of Physics and Astronomy, University of Leeds,  Leeds LS2 9JT, United Kingdom}
\author{Oriol Romero-Isart~\footnote{present address: Max-Planck-Institut f\"ur Quantenoptik,
Hans-Kopfermann-Strasse 1,
D-85748, Garching, Germany.}}
\affiliation{Departament de F\'isica, Universitat Aut\`onoma de Barcelona. Bellaterra, E-08193,  Catalonia, Spain}


\begin{abstract}
We show how the quantum Zeno effect can be exploited to control quantum many-body dynamics for quantum information and computation purposes.   
In particular, we consider a one dimensional array of three level systems interacting via a nearest-neighbour interaction. By encoding the qubit on two levels and using simple projective frequent measurements yielding the quantum Zeno effect, we demonstrate how to implement  a well defined quantum register, quantum state transfer on demand, universal two-qubit gates and two-qubit parity measurements. Thus, we argue that the main ingredients for universal quantum computation can be achieved in a spin chain with an {\em always-on} and {\em constant} many-body Hamiltonian.  We also show some possible modifications of the initially assumed dynamics in order to create maximally entangled qubit pairs and single qubit gates.
\end{abstract}

\vspace*{10pt}

\vspace*{3pt}

\maketitle


\section{Introduction}        
\label{sec:intro}
The dynamics that can be triggered in a quantum many-body system is in general very rich and complex. 
Many-body Hamiltonians, especially those related to strongly correlated systems, usually generate entanglement within the components of the many-body system which results in populating the exponentially large amount of orthogonal states available in the Hilbert space~\cite{amico08}.  Such dynamics is not only interesting from the fundamental point of view but also from its potential applications.  Some examples of the rich potentiality of quantum many-body systems are, among others, measurement-based quantum computation~\cite{raussendorf_one-way_2001}, where ground states of many-body Hamiltonians are used as the basic resource for the computation, adiabatic quantum computation~\cite{adiabatic}, which exploits the many-body dynamics of fine-tuned engineered interactions, and topological quantum computation~\cite{kitaev}, which hinges on the nonlocal properties of some many-body systems. 

In this work we aim at employing many-body dynamics for quantum information and computation purposes; by encoding qubits in the many-body system, the dynamics may be rendered on demand to move, transform, and entangle quantum information, hence, providing a powerful means to perform universal quantum computation. However, this implementation often requires a great amount of control of the dynamics. Such a control could be achieved, of course, if one would have the possibility to modify the interactions on demand. In this case, the engineering protocol would consist in a sequence of interaction tunings yielding a complex time dependent Hamiltonian. 

In this article, we show that this kind of control can also be achieved using an {\em always-on} many-body Hamiltonian. More specifically, we are interested in devising a scheme to move qubits and perform universal two-qubit gates within a many-body system in which the Hamiltonian is constant~\footnote{We also address possible extensions to our model, which provide extra functionality by introducing simple and reasonable modifications of the always-on dynamics.}. Indeed, we shall argue that this can be achieved even within a one-dimensional many-body system (spin chain hereafter). The key ingredient in order to control the dynamics will be the quantum Zeno effect~\cite{Zeno,Zenob}. The quantum Zeno effect prevents some particular subspaces from being populated and thus, renders an effective dynamics which in turn can take the role of the switching on and off of some interactions in the ideal case above mentioned.
To be more precise, the requirements in our scheme that renders a spin chain into a quantum information processing device are the following:
\begin{enumerate}
	\item[\em a)] A spin-1 chain with an \emph{always-on} nearest-neighbour exchange interaction.
	\item[\em b)] Frequent projective measurements on each site, that discriminate one level $\ket0$ (used as a {\em vacuum}) against the other two $\{\ket\up,\ket\down\}$ (encoding a {\em qubit}).
\end{enumerate}


\begin{figure*}
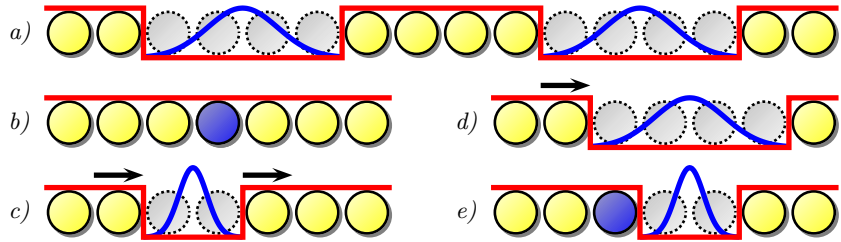

\begin{center}
\psset{unit=.66cm}
\pspicture(16,1)
\rput(-.5,.5){\em a)}
\groc{.5,.5}
\groc{1.5,.5}
\multido{\r=2.5+1,\i=-1+1}{4}{\none{\r,.5}}
\groc{6.5,.5}
\groc{7.5,.5}
\groc{8.5,.5}
\groc{9.5,.5}
\multido{\r=10.5+1,\i=-1+1}{4}{\none{\r,.5}}
\groc{14.5,.5}
\groc{15.5,.5}
\rput(4,0){\psplot[linecolor=blue,linewidth=2pt,plotstyle=curve]{-2}{2}{2.7 -1 x 2 exp mul exp} }
\rput(12,0){\psplot[linecolor=blue,linewidth=2pt,plotstyle=curve]{-2}{2}{2.7 -1 x 2 exp mul exp} }
\psline[linecolor=red,linewidth=2pt](0,1)(2,1)(2,0)(6,0)(6,1)(10,1)(10,0)(14,0)(14,1)(16,1)
\endpspicture
\vspace{.5cm}

\pspicture(16,1)
\rput(-.5,.5){\em b)}
\multido{\r=.5+1,\i=-1+1}{7}{\groc{\r,.5}}
\qubit{3.5,.5}
\psline[linecolor=red,linewidth=2pt](0,1)(7,1)

\rput(8.5,.5){\em d)}
\groc{9.5,.5}
\groc{10.5,.5}
\multido{\r=11.5+1,\i=-1+1}{4}{\none{\r,.5}}
\groc{15.5,.5}
\rput(13,0){\psplot[linecolor=blue,linewidth=2pt,plotstyle=curve]{-2}{2}{2.7 -1 x 2 exp mul exp} }
\psline[linecolor=red,linewidth=2pt](9,1)(11,1)(11,0)(15,0)(15,1)(16,1)
\psline[linewidth=2pt]{->}(10,1.25)(11,1.25)
\endpspicture
\vspace{.5cm}

\pspicture(16,1)
\rput(-.5,.5){\em c)}
\groc{.5,.5}
\groc{1.5,.5}
\none{2.5,.5}
\none{3.5,.5}
\multido{\r=4.5+1,\i=-1+1}{3}{\groc{\r,.5}}
\rput(3,0){\psplot[linecolor=blue,linewidth=2pt,plotstyle=curve]{-1}{1}{1.4 500 -1 x 2 exp mul exp mul} }
\psline[linecolor=red,linewidth=2pt](0,1)(2,1)(2,0)(4,0)(4,1)(7,1)
\psline[linewidth=2pt]{->}(1,1.25)(2,1.25)
\psline[linewidth=2pt]{->}(4,1.25)(5,1.25)

\rput(8.5,.5){\em e)}
\groc{9.5,.5}
\groc{10.5,.5}
\qubit{11.5,.5}
\none{12.5,.5}
\none{13.5,.5}
\groc{14.5,.5}
\groc{15.5,.5}
\rput(13,0){\psplot[linecolor=blue,linewidth=2pt,plotstyle=curve]{-1}{1}{1.4 500 -1 x 2 exp mul exp mul} }
\psline[linecolor=red,linewidth=2pt](9,1)(12,1)(12,0)(14,0)(14,1)(16,1)
\endpspicture
\vspace{.35cm}
\end{center}
\caption{ Red bars represent measured sites (upper) and unmeasured sites (lower), yellow sites are atoms yielding outcome $\ket0\bra0$ and blue sites yield outcome $\openone-\ket0\bra0$. Grey sites are not measured. {\em a)} 2 Blocks encoding 1 qubit each. {\em b)} A qubit frozen into site $k$. {\em c)} State transfer using a size-2 block. {\em d)} State transfer by the compression scheme. {\em e)}~Two-qubit gate or a parity measurement using a fixed qubit at the block boundary and a free qubit inside the block.}
\label{fig:sketch}
\end{figure*}

If these requirements are met, a quantum information processing toolbox with the following features can be achieved:
\begin{enumerate} 
	\item One can easily implement a quantum register where well-defined qubits can be identified and stored.
	\item Qubits can be transferred {\em perfectly}  from any site to any other, on demand. The transfer can be halted or modified at any time. Qubits can be delivered at a rate which is independent of the distance. 
	\item Universal two qubit gates of the form $\sqrt[\alpha]{SWAP}$ with arbitrary $\alpha$ can be implemented by a repeat-until-success scheme.
	\item Two-qubit projective parity measurements can be implemented by a repeat-until-success scheme.  
\end{enumerate}
We also show possible extensions to the requirements stated above, in order to incorporate the following:
\begin{enumerate}
	\setcounter{enumi}{4}
	\item Spontaneous two-qubit pair creation in a maximally entangled state $\ket{\up\downarrow}+\ket{\down\uparrow}$.
	\item Single qubit gates implementation without local perturbations but only local measurements in single sites of the chain.
\end{enumerate} 

The paper is organized as follows. In Section~\ref{sec:preliminaries} we define more precisely the system, dynamics and measurements required in our scheme, as well as introducing some notational conventions. In Section~\ref{sec:qregister}, we explain how to implement a quantum register by means of the Zeno effect. Section~\ref{sec:transfer} is devoted to show various state transfer procedures. Section~\ref{sec:2qbdynamics} explores the possibilities that the dynamics of two qubits offer, showing how two-qubit gates [Subsection~\ref{sec:2qbgates}] and projective parity measurements [Subsection~\ref{sec:2qbparity}] can be implemented. Finally, Section~\ref{sec:extensions} explores some extensions to our list of assumptions that allow to incorporate maximally entangled pair creation, and single qubit gates. The work concludes with some final remarks.

\section{Preliminaries} 
\label{sec:preliminaries}
Let us begin the discussion introducing our many-body system: a one dimensional (1D) array of $n$ 3-level systems (hereafter regarded as a spin-1 chain), coupled by a  swap interaction of the form
\begin{equation} \label{eq:hamiltonian}
	   H=J\sum_{\langle i,j\rangle}    W_{ij},
\end{equation}
where $J$ is the coupling strength, $\langle\cdot,\cdot\rangle$ refers to nearest neighbours and $   W_{ij}$ is the ${SU}(3)$ swap operator between the states in site $i$ and $j$. This Hamiltonian is a particular case, called \emph{Uimin-Lai-Sutherland} model~\cite{ULS,ULSb,ULSc}, of a spin-$1$ chain governed by the most general isotropic Hamiltonian with nearest neighbor interactions
\begin{equation}
	\label{eq:GenHeisHam}
	  H=\tilde J\sum_i (\cos \theta~\vec S_i \vec S_{i+1}+\sin \theta (\vec S_i \vec S_{i+1})^2),
\end{equation}
where $  S^{x,y,z}_i$ are the usual spin-$1$ operators at site $i$. 
For $\theta=\pi/4$ Eq.~\eref{eq:hamiltonian} is recovered, up to irrelevant constant terms, where $J=\tilde J/\sqrt2$. This Hamiltonian, as will be shown, preserves the quantum information and naturally generates a quantum walk which can be driven using measurements.

 Spin levels will be denoted by $\{\ket{\up},\ket 0,\ket{\down}\}$. A qubit can be encoded using the levels  $\{\ket\up, \ket \down\}$. The absence of qubit will be encoded with the state $\ket{0}$, called \emph{vacuum}. Note that this choice of states is completely arbitrary and for the main part of the paper, they can be regarded simply as \emph{labels}. Only in Section~\ref{sec:extensions} it will matter what physical state is chosen to represent the vacuum.
 
  We can introduce the number operator in each one of the levels, $   N_m=\sum_i   P_m^{i}$, where $   P_m^{i}=\ket{m}_i\bra{m}$ is the projector onto state $\ket{m}$ in site $i$ ($m=\uparrow, 0,\downarrow$). The number operators $\{  N_m\}$ are conserved quantities, $[  H,   N_m]=0$. The number operator $  N=n\openone-  N_{0}=  N_{\downarrow}+  N_{\uparrow}$ counts the \emph{number of qubits} encoded in the chain. From now on we will specialize to the 1-qubit subspace ($\mathcal H_1$), namely, states with quantum number $N=1$. Separable one-qubit states can be written as $\ket{\psi;\bf k}$, meaning that the qubit $\ket{\psi}=\alpha \ket \up+\beta \ket \down$ is encoded in site $k$, namely
\begin{equation}
	\label{eq:1exc}
	\ket{\psi;{\bf k}}=\ket{0}_1\otimes\cdots\otimes(\alpha\ket{\up}_k+\beta\ket{\down}_k)\otimes\cdots\otimes\ket{0}_n.
\end{equation}
%
These states span the 1-qubit subspace $\mathcal H_1$, which can be naturally factorized as $\mathcal H_1\cong\mathcal Q \otimes \mathcal L=\textrm{span}\{\ket{\up;{\bf k}},\ket{\down;{\bf k}}, {\bf k}=1,\ldots,n\}$, where $\mathcal Q=\mathbb C^2$ represents the qubit Hilbert space and $\mathcal L=\mathbb C^n$ is the Hilbert space for the position degree of freedom. With this factorization the Hamiltonian acts trivially on $\mathcal Q$, \emph{i.e.}, $  H=\openone\otimes   \hat H_\mathcal{L}$, and separable one-particle states naturally factor into $\ket{\psi;\bf k}=\ket{\psi}\otimes\ket{\bf{k}}$. Thus, hereafter $\ket{\bf{k}}$ indicates that the qubit is at site $k$. We will use the hat to denote operators acting \emph{only} on the position space $\mathcal L$.

Let us now discuss the effect of continuous measurements on some sites. The dynamics under continuous measurement (represented by projectors $\{  E_{\chi}: \sum_{\chi}   E_{\chi}=\openone\}$) can be described by the effective Hamiltonian 
\begin{equation}
	H_{\eff}=  E_{\chi}   H   E_{\chi},
\end{equation}
where $\chi$ is the outcome of the measurement. This is usually called {\em quantum Zeno dynamics}~\cite{facchi_quantum_2007}. The QZE will allow to tailor the evolution of our system on demand. The measurement that allows to perform the desired tasks is
\begin{equation}
	\label{eq:kmeasurement}
	  E_\pout^k=  P_{0}^k \qquad   E_\pin^k=\openone-  P_{0}^k,
\end{equation}
which corresponds to determining whether site $k$ contains a qubit ($\chi=\pin$) or not ($\chi=\pout$). It is understood that these act trivially ($\openone$) on the remaining sites. The representation of the $E_\pin^k, E_\pout^k$ operators in $\mathcal H_1$ can be obtained from
\begin{equation}
	\bra{x';{\bf k'}}E_\pout^k\ket{x'';{\bf k''}}=\delta_{x'x''}\delta_{k'k''}\left(1-\delta_{kk'}\right),
\end{equation} 
where $x=\uparrow,\downarrow$ and $k$ takes values on all sites. These matrix elements impliy that
\begin{eqnarray}\nonumber
	E_\pout^k&=&\sum_{x'x''}\sum_{k'k''}\delta_{x'x''}\delta_{k'k''}\left(1-\delta_{kk'}\right)\ket{x';{\bf k'}}  \bra{x'';{\bf k''}}\\ \nonumber 
	&=&\openone_{\mathcal Q}\otimes\sum_{k'}\left(1-\delta_{kk'}\right)\ket{{\bf k'}}  \bra{{\bf k'}}\\
	&=&{\openone_{\mathcal Q}}\otimes\left(
	\openone_{\mathcal L}-\ket{{\bf k}}\bra{{\bf k}}\right).
\end{eqnarray}
Here we have used $\sum_{x'x''}\delta_{x'x''}\ket{x'}\bra{x''}=\openone_{\mathcal Q}$ and $\sum_{k}\ket{{\bf k}}\bra{{\bf k}}=\openone_{\mathcal L}$. From this, one obtains the representation of $E_\pin^k$ since
\begin{equation}
	E_\pin^k=\openone_{\mathcal Q}\otimes\openone_{\mathcal L}-{\openone_{\mathcal Q}}\otimes\left(\openone_{\mathcal L}-\ket{{\bf k}}\bra{{\bf k}}\right)={\openone_{\mathcal Q}}\otimes\ket{{\bf k}}\bra{{\bf k}}
\end{equation}
 From now on we ommit the subscripts $\mathcal Q$ and $\mathcal L$ when no confusion arises. Since both measurement projectors act trivially on the qubit space $\mathcal Q$ it is convenient to introduce the position projectors by observing that $E_\pin^k=\openone\otimes  \hat {\mathcal E}_\pin^k$ ($E_\pout^k=\openone\otimes\hat\mathcal E_\pout^k$), where $ \hat {\mathcal E}_\pin^k~(\hat {\mathcal E}_\pout^k)$ acts on $\mathcal L$ and is given by $ \hat {\mathcal E}_\pin^k=\ket{\bf{k}}\bra{\bf{k}}$ ($\hat\mathcal E_\pout^k=\openone-\ket{\bf k}\bra{\bf k}$). Thus, the action of this measurement is effectively questioning whether the qubit is located at position $k$ or not, without revealing the actual state of the qubit. The effect of measuring more than one site at a time (sites $K=\{k_1,\ldots,k_r\}$) can be described by the projectors
\begin{equation}
 	 \hat {\mathcal E}_0=\openone-\sum_{{\bf k}\in K}\ket{{\bf k}}\bra{{\bf k}},\quad \hat {\mathcal E}_k=\ket{{\bf k}}\bra{{\bf k}},\quad k\in K.
\end{equation}
With this notation, measurement of the $  P_0$ observable on all sites of the block implements the observable $\pos=\sum_k k\,\ket{ {\bf k}}\bra{ {\bf k}}=\sum_k k\,\hat\mathcal E_k$ on space $\mathcal L$.
\section{Quantum register}
\label{sec:qregister}

We now analyze the dynamics of the spin chain in the 1-qubit subspace, and the effect of the continuous measurements. We will show how different parts of the chain can be decoupled, giving rise to the notion of \emph{blocks}, which can encode one qubit each, turning the spin chain into a quantum register.

We begin by showing the effect a continuous measurement $\{E_\pin^k,E_\pout^k\}$  with outcome ``$\pout$'' has on the effective Hamiltonian. The effective dynamics will be given by the operators $P_0^k W_{ij} P_0^k$, because the measured site is encountered in the $\ket0$ state. These operators read
\begin{equation}\label{eq:effectiveW}
	P_0^k W_{ij} P_0^k=\left\{ \begin{array}{ll}
		W_{ij}\otimes \ket0_k\bra0&k\neq i,j\\
		\ket0_i\bra0\otimes\ket0_j\bra0&k=i \textrm{~or~} k=j
	\end{array} \right.
\end{equation}
where the identity is assumed in all the other sites. This is easily proven by writing $P_0^kW_{ij}P_0^k=\sum_{\mu\nu}\ket0_k\bra0\big(\ket\nu_i\bra\mu\otimes\ket\mu_j\bra\nu\big)\ket0_k\bra0$, where $\mu,\nu$ run over all spin-1 states. Whenever $k$ equals $i$ or $j$, the sum $\sum_{\mu\nu}$ immediately collapses into $\ket0_i\bra0\otimes\ket0_j\bra0$. If $k\neq i,j$ no further simplification is possible.

A key observation is that if an $\{E_\pin,E_\pout\}$ continuous measurement is performed on a site, say $k$, (or set of neighbouring sites) and the outcome ``$\pout$'' is obtained, the two sides of this site are decoupled. This can be seen by realizing that the only terms in the Hamiltonian \eref{eq:hamiltonian} that couple site $k$ to its left and right neighbours are $  W_{k-1,k}+  W_{k,k+1}$. Taking into account expression \eref{eq:effectiveW}, these terms effectively become 
\begin{equation}
	\ket{0}_{k-1}\bra{0}\otimes\ket{0}_k\bra{0}\otimes\openone_{k+1}+\openone_{k-1}\otimes\ket{0}_k\bra{0}\otimes\ket{0}_{k+1}\bra{0}
\end{equation}
where subindices refer to chain sites. Adding all the remaining terms in Eq.~\eref{eq:hamiltonian} the effective Hamiltonian can be written as 
\begin{equation}
	  H_\eff=J(H_A\otimes\ket{0}_k\bra{0}\otimes\openone_B+\openone_A\otimes\ket{0}_k\bra{0}\otimes H_B),
\end{equation}
where $H_A=P_0^{k-1}+\sum_{i=1}^{k-2}W_{i,i+1}$, and $H_A=P_0^{k+1}+\sum_{i=k+1}^{n-1}W_{i,i+1}$. Since these operators commute, the time evolution factors into
\begin{equation}
  U(t)=  U_A(t)\otimes \openone_k \otimes   U_B(t).
\end{equation}
Hence, the chain is effectively divided into two parts $A$ and $B$ which evolve independently, separated by a site in the $\ket{0}$ state. More generally, the chain can be split into \emph{blocks} by performing continuous measurements in various points along the chain [see Fig.~\ref{fig:sketch}\emph{a}]. Each block can be thought of as an individual chain with its own dynamics, given by the Hamiltonian 
\begin{equation}
	  H_\eff=J(\ket 0_l\bra0+\sum_{i=l}^{r-1}   W_{i,i+1}+\ket0_r\bra0),
\end{equation}
where the $l$ and $r$ subscripts represent the outermost \emph{left} and \emph{right} sites of the block, respectively, and the $\ket0_{l,r}\bra0$ terms arise from the boundary interaction under Zeno dynamics. For simplicity we have considered one site to split the chain, but the argument is also valid if several contiguous sites are measured. Note that a similar approach has been taken in a three-site system to perform two qubit gates \cite{benjamin}. 

In the following we will focus on a single block of size $n$, with sites labelled from $1$ to $n$. The effective Hamiltonian 
reads $  H_\eff=\openone\otimes   \hat H_\ch$, with $\hat H_\ch$ acting on $\mathcal L$, given by (in the basis $\{\ket{{\bf 1}},\ldots,\ket{ {\bf n}}\}$)
\begin{equation}
\label{eq:HeffL}
 	  \hat H_\ch=J\sum_{k=1}^{n-1} \ket{{\bf k+1}}\bra{{\bf k}}+\ket{{\bf k}}\bra{{\bf k+1}}
\end{equation}
plus an irrelevant constant term.

Assume the block is initially prepared in the state $\ket{\psi;{\bf k}}$ and let $  \rho$ represent its corresponding density operator, $  \rho=\ket{\psi;{\bf k}}\bra{\psi;{\bf k}}$. The natural evolution under the Hamiltonian $  H_\ch$ corresponds to a delocalization of the qubit, by a continuous time quantum walk~\cite{quantum_walks,quantum_walksb} with Dirichlet boundary conditions $\bra{{\bf 0}}
\rho\ket{{\bf 0}}=\bra{{\bf n+1}}\rho\ket{{\bf n+1}}=0$, while preserving its internal state. The position of the qubit can be localized on-demand by simply measuring $\pos$. Moreover, if $\pos$ is performed continuously, the time evolution is frozen and delocalization is prevented [see Fig~\ref{fig:sketch}\emph{b}]. In fact, when this is the case, the block is reduced to a single site, e.g., that encoding the qubit. Finding the qubit in site $k$ corresponds to the projector $ \hat {\mathcal E}_k$, thus the density operator reads $\rho=\rho_{\mathcal Q}\otimes\hat \mathcal E_k$, where $\rho_{\mathcal Q}$ is the qubit state and trivially $[\hat {\mathcal E}_k\hat H_\ch\hat {\mathcal E}_k, \hat {\mathcal E}_k]=0$, thus $[H_\eff,\rho]=0$. The quantum Zeno dynamics generated by the continuous measurement of $\pos$ prevents $  \rho$ from evolving outside the space generated by $\hat  {\mathcal E}_k$. 

In summary, the spin chain can be split into blocks, each one containing a delocalized qubit, that can be instantly and continuously localized by measuring all sites of the block in which it is encoded.
\section{State transfer} 
\label{sec:transfer}

\begin{figure*}[t]
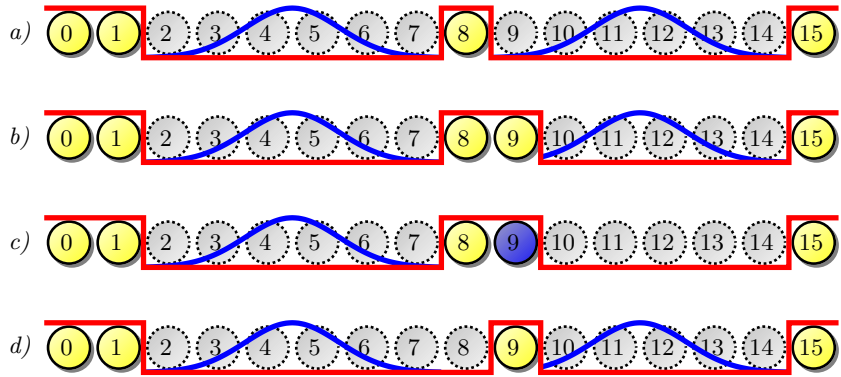

\begin{center}
\psset{unit=.66cm}

\pspicture(16,1)
\rput(-.5,.5){\em a)}
\groc{.5,.5}
\groc{1.5,.5}
\multido{\r=2.5+1,\i=-1+1}{6}{\none{\r,.5}}
\groc{8.5,.5}
\multido{\r=9.5+1,\i=-1+1}{6}{\none{\r,.5}}
\groc{15.5,.5}
\multido{\n=0+1}{16}{\rput(\n,.5){~~~~\,\,\,\n}}

\rput(5,0){\psplot[linecolor=blue,linewidth=2pt,plotstyle=curve]{-3}{3}{2.7 -1 .75 x mul 2 exp mul exp} }
\rput(12,0){\psplot[linecolor=blue,linewidth=2pt,plotstyle=curve]{-3}{3}{2.7 -1 .75 x mul 2 exp mul exp} }
\psline[linecolor=red,linewidth=2pt](0,1)(2,1)(2,0)(8,0)(8,1)(9,1)(9,0)(15,0)(15,1)(16,1)
\endpspicture
\vspace{.7cm}

\pspicture(16,1)
\rput(-.5,.5){\em b)}
\groc{.5,.5}
\groc{1.5,.5}
\multido{\r=2.5+1,\i=-1+1}{6}{\none{\r,.5}}
\groc{8.5,.5}
\groc{9.5,.5}
\multido{\r=10.5+1,\i=-1+1}{5}{\none{\r,.5}}
\groc{15.5,.5}
\multido{\n=0+1}{16}{\rput(\n,.5){~~~~\,\,\,\n}}

\rput(5,0){\psplot[linecolor=blue,linewidth=2pt,plotstyle=curve]{-3}{3}{2.7 -1 .75 x mul 2 exp mul exp} }
\rput(12,0){\psplot[linecolor=blue,linewidth=2pt,plotstyle=curve]{-2}{3}{2.7 -1 .75 x mul 2 exp mul exp} }
\psline[linecolor=red,linewidth=2pt](0,1)(2,1)(2,0)(8,0)(8,1)(10,1)(10,0)(15,0)(15,1)(16,1)
\endpspicture
\vspace{.7cm}

\pspicture(16,1)
\rput(-.5,.5){\em c)}
\groc{.5,.5}
\groc{1.5,.5}
\multido{\r=2.5+1,\i=-1+1}{6}{\none{\r,.5}}
\groc{8.5,.5}
\qubit{9.5,.5}

\multido{\r=10.5+1,\i=-1+1}{5}{\none{\r,.5}}
\groc{15.5,.5}
\multido{\n=0+1}{16}{\rput(\n,.5){~~~~\,\,\,\n}}

\rput(5,0){\psplot[linecolor=blue,linewidth=2pt,plotstyle=curve]{-3}{3}{2.7 -1 .75 x mul 2 exp mul exp} }
\psline[linecolor=red,linewidth=2pt](0,1)(2,1)(2,0)(8,0)(8,1)(10,1)(10,0)(15,0)(15,1)(16,1)
\endpspicture
\vspace{.7cm}

\pspicture(16,1)
\rput(-.5,.5){\em d)}
\groc{.5,.5}
\groc{1.5,.5}
\multido{\r=2.5+1,\i=-1+1}{7}{\none{\r,.5}}
\groc{9.5,.5}
\multido{\r=10.5+1,\i=-1+1}{5}{\none{\r,.5}}
\groc{15.5,.5}
\multido{\n=0+1}{16}{\rput(\n,.5){~~~~\,\,\,\n}}
\rput(5,0){\psplot[linecolor=blue,linewidth=2pt,plotstyle=curve]{-3}{3}{2.7 -1 .75 x mul 2 exp mul exp} }
\rput(12,0){\psplot[linecolor=blue,linewidth=2pt,plotstyle=curve]{-2}{3}{2.7 -1 .75 x mul 2 exp mul exp} }
\psline[linecolor=red,linewidth=2pt](0,1)(2,1)(2,0)(9,0)(9,1)(10,1)(10,0)(15,0)(15,1)(16,1)
\endpspicture
\vspace{.3cm}\\
\end{center}
\caption{
{\em a)} 2 Blocks encoding 1 qubit each. 
{\em b)} The left block boundary of the rightmost block is shifted by one site by continuously measuring site 9. If no qubit is encountered in site 9, the shifting is successful.
{\em c)} If the qubit is encountered in site 9 the shifting unsuccessful. Measuring site 8 prevents the qubit from interacting with the qubit encoded in the block on the left. Going back to measurement scheme (\emph{a}) allows to start the procedure anew. 
{\em d)} If the shifting is successful, one can \emph{release} site 8 for the next block.
}
\label{fig:blocks}
\end{figure*}

The notion of blocks will be central throughout the remaining of the paper. In this section we show that shifting and resizing blocks provides a simple way to perform perfect state transfer with a large degree of versatility. Shifting a block (or its boundary) corresponds to changing the set of sites being measured [Fig.~\ref{fig:blocks}]. For example, suppose a block from site 9 to site $14$ is defined by measuring sites $K=\{8,15\}$ and one wants to shift the left boundary ($8$) by one site to the right [Fig.~\ref{fig:blocks}\emph{a}]. This is accomplished by, at any given time, starting to measure also site $9$, so that $K$ becomes $\{8,9,15\}$ [Fig.~\ref{fig:blocks}\emph{b}]. It is necessary to keep measuring site $8$ for a small amount of time, in case the qubit would be encountered in site $9$, to prevent it from interacting with the block to the left (which may contain another qubit). In the event that the qubit is found in site $9$ [Fig.~\ref{fig:blocks}\emph{c}], one would go back to the initial configuration of measurements [Fig.~\ref{fig:blocks}\emph{a}], and after a given time (comparable to the timescale $J^{-1}$) the shifting is attempted anew. This procedure trivially extends to shifting and resizing whole blocks.

In this scheme, the qubit evolves as a continuous time quantum walk driven by the QZE, namely, by measuring the appropriate sites the space where the walk takes place can be dynamically changed.
Before entering the discussion in more detail, we want to point out that spin-1 chains have previously been studied in the context of state transfer~\cite{spin1chains,spin1chainsb,spin1chainsc,spin1chainsd,spin1chainse}. Although the measurement in Eq.~\eref{eq:kmeasurement} was used in the double rail encoding to check the arrival of the qubit without destroying its state~\cite{spin1chainsb}, the possibility of enhancing the performance by the QZE was not considered.
\begin{figure}[b]
\begin{center}
\includegraphics[width=.45\linewidth]{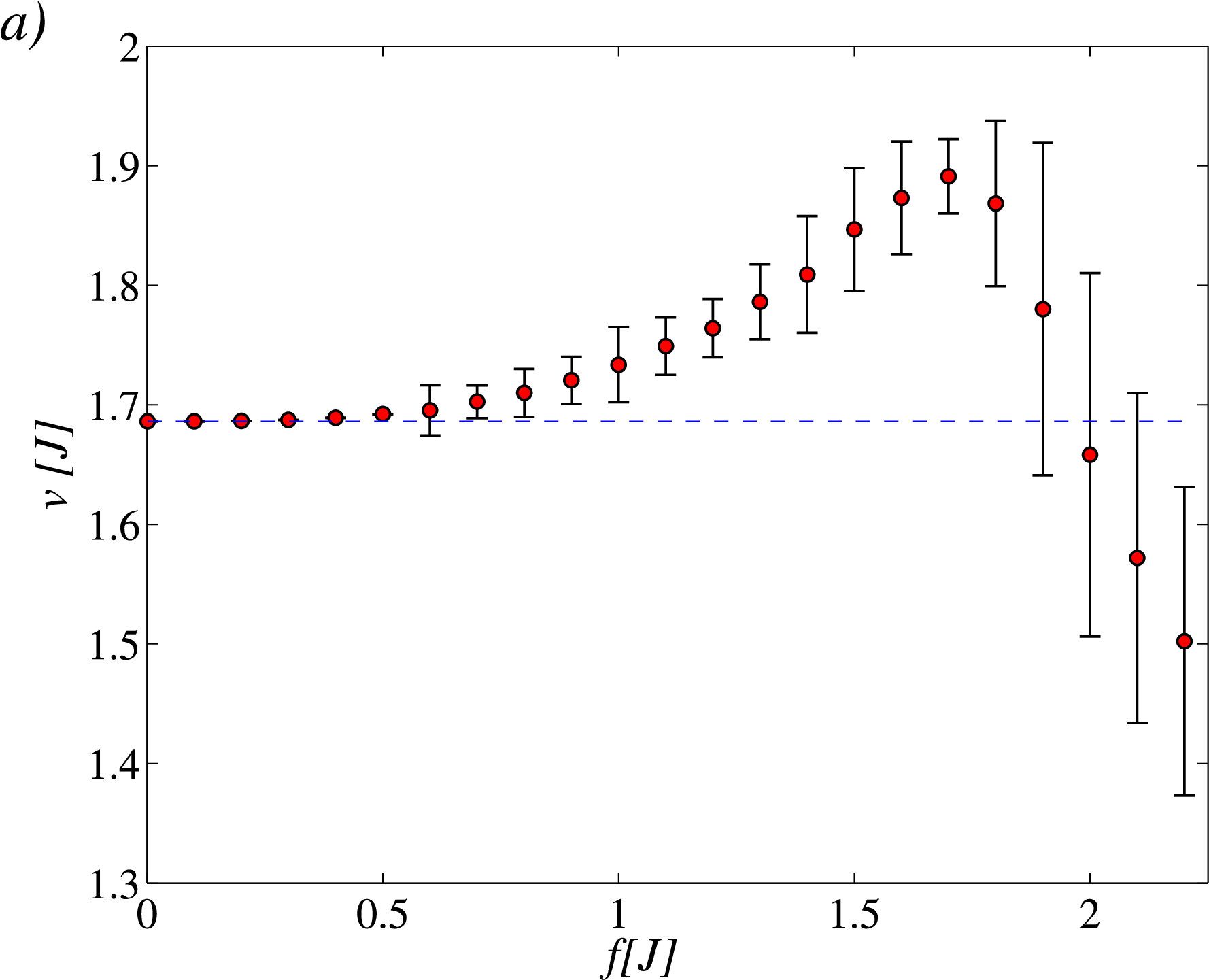}
\includegraphics[width=.5\linewidth]{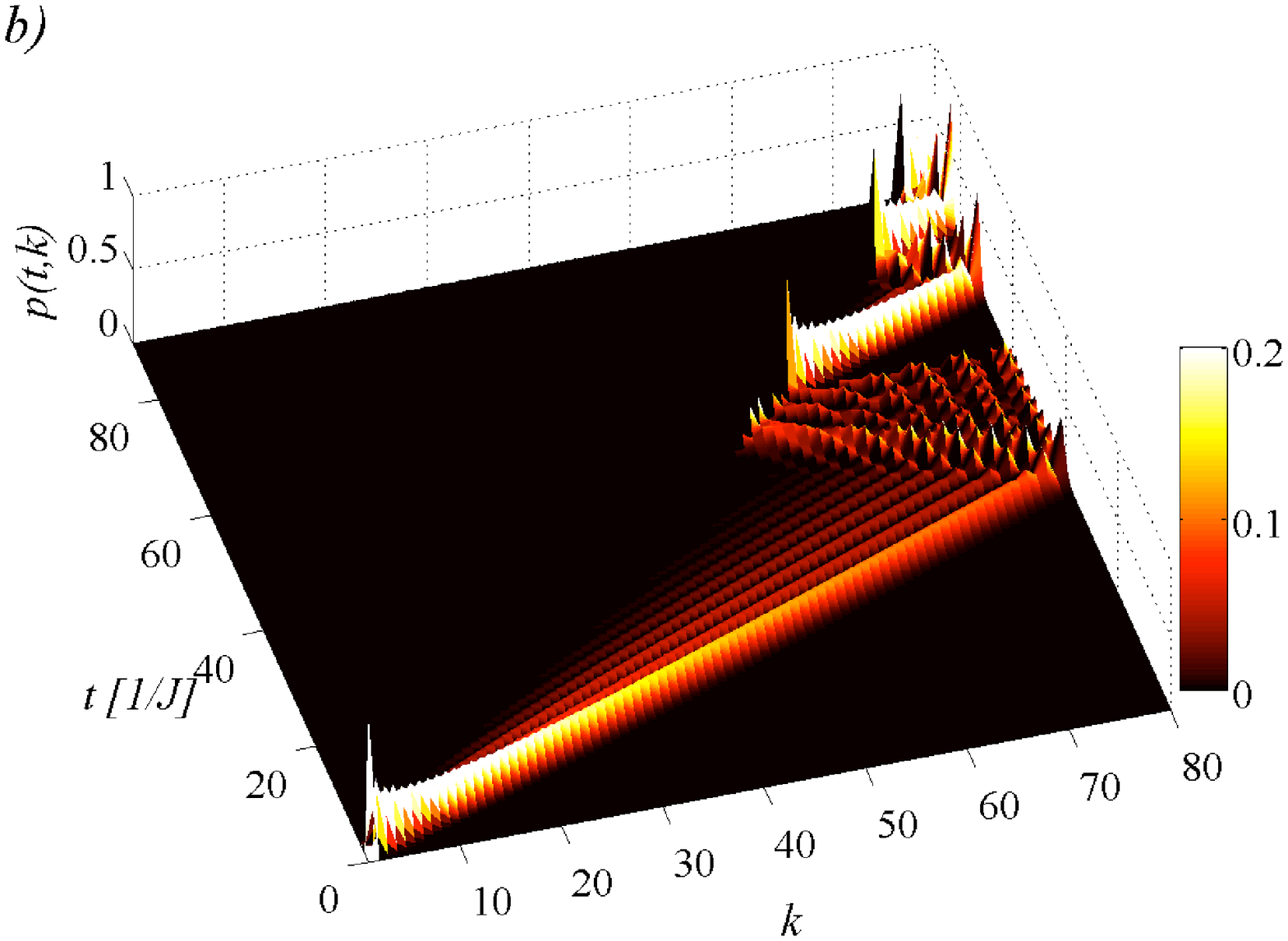} 
\end{center}
\caption{
\emph{a)} Transfer speed ($d\langle \pos\rangle/dt$) for scheme 3 as a function of the rate at which the boundary is shifted ($f=\Delta t^{-1}$). Dashed line represents free evolution, (scheme 2) $f=0$. Notice that for frequencies at about $J\lesssim f\lesssim 2J$ one observes the anti-Zeno effect. For larger frequencies ($f>2J$) the QZE takes over. A time interval of $4\Delta t$ is left after the qubit is found in the boundary to avoid the QZE and enhance anti-Zeno. Error bars represent fluctuations of the transfer protocol. \emph{b)} Density profile $p(t,k)=|\bra{{\bf k}}   \hat U(t) \ket{{\bf 1}} |^2$ for the compressing scheme, $f=J$ and $n=80$. 
}
\label{fig:anti-zeno}
\end{figure}

As mentioned above, perfect state transfer is achieved by shifting the boundaries of the block where the qubit is encoded, and there are several ways of doing so. 

\begin{itemize}
\item {\em Scheme 1}, ``SWAP concatenation'' [see Fig~\ref{fig:sketch}\emph{c}]: by encoding the qubit in a block of size 2, the effective Hamiltonian in $\mathcal L$ reads $  \hat H_\ch=J(\ket{{\bf 1}}\bra{{\bf 2}}+\ket{{\bf 2}}\bra{{\bf 1}})$. Hence, the time evolution is $\hat U(t)=\exp(-it  H_\ch)$, which for $\Delta t=\pi/(2J)$ performs an X gate $\hat U(\Delta t)=  \hat X_{1,2}$ (up to global phases) in localization space. Then, the block is shifted by one site and the process is started over (Fig.~\ref{fig:blocks}). This procedure is deterministic and provides a speed of $2J/\pi\simeq0.64J$ sites per unit time. 

An anti-Zeno effect~\cite{balachandran_quantum_2000} may be invoked by reducing $\Delta t$. In this case the block boundaries are shifted before the qubit is deterministically transferred and the total time for transferring $n$ sites is $T=\Delta t(n+q)$ where $q$ is the number of times the swap \emph{fails} and the procedure needs to be repeated. The failure probability is $p=\cos^2J\Delta t$. The probability for occurring $q$ failures spread over $n+q$ trials is
\begin{equation}
	p(q)=\left(\begin{array}{c}n+q \\q\end{array}\right)p^q(1-p)^n,
\end{equation}
and the mean time, averaged over all possible number of failures, is
\begin{equation}
\langle T\rangle=\Delta t\sum_{q=0}^\infty(n+q) p(q)=\Delta t\frac{1+n}{(1-p)^2},
\end{equation}
which for large transfer lengths ($n\gg 1$) is
\begin{equation}
	\langle T\rangle\simeq \frac{\Delta t n}{(1-p)^2}=\Delta t \csc^4 J\Delta t.
\end{equation}
Optimizing $\langle T\rangle$ yields $\Delta t=1.39/J$, giving an average speed increase of about 5\%. 

Notice that this scheme only employs a finite number of sites  at once, and hence it allows for several qubits to simultaneously occupy the chain, distributed among their corresponding blocks. This is not a significant feature if only one qubit needs to be transferred. However, in most situations one is interested in a high qubit delivery rate, rather than a short time delay between sender and receiver. With this scheme, the time delay increases linearly with the distance, but the delivery rate (qubits sent/received per unit time) remains constant. This is an important requirement for large scale quantum communication.

\item {\em Scheme 2}, imaging: the qubit is initially near the left-end of a block and is left to evolve freely. After a long time $\Delta T$ a measurement $\pos$ is performed and the left boundary of the block is shifted next to the qubit. The procedure is repeated until the block has the desired size and position. The transfer speed with this procedure, assuming that the transfer distance is large ($n\gg1$), and thus the block is always much larger that the width of the wave packet, corresponds to the free propagation speed of a continuous-time quantum walk with one boundary. This can be numerically evaluated from~\cite{bessen}
\begin{eqnarray}\nonumber
	\langle\pos\rangle&=&\sum_{x=1}^\infty x\left(\frac{Jx}{t}{\bf J}_x(2t/J)\right)^2,\\
	v&=&\lim_{t\rightarrow\infty}\langle\pos\rangle/t\simeq 1.69J.
\end{eqnarray}
(this value corresponds to the dashed line in Fig.~\ref{fig:anti-zeno}\emph{a}). ${\bf J}_\nu(z)$ is the Bessel function of the first kind~\cite{abram}.

\item {\em Scheme 3}, ``compressing"  [see Fig~\ref{fig:sketch}\emph{d}]: assume the qubit is encoded in site $1$ (left) which needs to be transferred to site $n$ (right). Initially one sets a block from site $1$ to site $n$ and shifts the left boundary by one site [see Fig.~\ref{fig:blocks}] at time intervals $\Delta t$, thus compressing the qubit to the desired site  [see Fig. \ref{fig:anti-zeno}\emph{b}]. 
If the measurement performed by the block boundary encounters the qubit, the block boundary steps back and halts for a given time.
 For large $\Delta t$ the qubit will propagate freely along the chain. For small $\Delta t$ the QZE will be invoked, thus hindering the qubit from propagating. At some optimal $\Delta t$ the anti-Zeno effect will push the qubit forward [See Fig. \ref{fig:anti-zeno}\emph{a}], providing a speedup of about 12\% with respect to the free propagation speed. A constant qubit-rate for schemes 2 \& 3 could also be achieved by allowing to move the right boundary of each block, as in scheme 1. 
\end{itemize}
These three schemes exploit the fact that the block boundaries act as potential barriers and the qubit propagates as a free particle in the lattice. Combinations of these schemes can provide a variety of performance \emph{vs.} control tradeoffs and thus can be adapted to meet different demands depending on the particular requirements and limitations of a given physical implementation. It is worth noticing that schemes 2 \& 3, which use large blocks perform significantly faster than scheme 1. The reason for this is that in scheme 1, the qubit is prevented from advancing too fast since it is constrained to a size-2 block. In all cases, the Anti-Zeno effect provides a noticeable increase in performance.

Summarizing and comparing to novel proposals for quantum communication in spin chains \cite{bose_overview_2007}, our proposal has the following particular properties:
\begin{itemize}
\item[{\em i)}] The qubit-rate does not decrease with the length of the chain (as we are not restricted to populate the chain with only one qubit at any given time), and the anti-Zeno effect~\cite{balachandran_quantum_2000} can be exploited to improve the transfer speed without affecting the overall qubit-rate. 
\item[{\em ii)}] The transfer can be modified during the process, that is, it can be stopped or the direction can be changed.
\item[{\em iii)}] Different schemes can be designed offering a variety of control-qubit-rate tradeoffs, which render this proposal highly versatile. 
\end{itemize}


\section{Two-qubit dynamics: Gates and Measurements}
\label{sec:2qbdynamics}
So far we have concentrated on single qubit dynamics. It was shown in Sec. \ref{sec:qregister} that the effective Zeno dynamics of a single qubit in a block only affects the position degree of freedom. This leads to a simple description of the time evolution, which decouples the qubit state from the position degree of freedom. Moreover, the dynamics for the qubit state becomes trivial, and hence the single qubit state is preserved. The blocks considered so far assumed that the boundaries of the block contain no qubit, \emph{i.e.}~the boundary is always set to the $\ket0$ state.

\begin{figure}
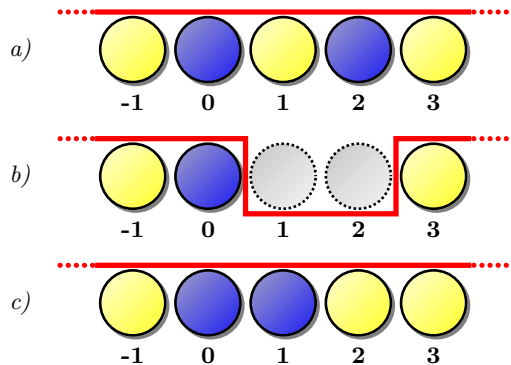

\begin{center}
\pspicture(5,1)
\psset{linewidth=2pt}
\rput(-1,.5){\em a)}
\psline[linestyle=dotted,linecolor=red,dotsep=1pt](-.5,1)(0,1)
\psline[linecolor=red](0,1)(2,1)(5,1)
\psline[linestyle=dotted,linecolor=red,dotsep=1pt](5,1)(5.5,1)
\psset{linewidth=1pt}

\groc{.5,.5}
\qubit{1.5,.5}
\groc{2.5,.5}
\qubit{3.5,.5}
\groc{4.5,.5}

\multido{\r=.5+1,\i=-1+1}{5}{\rput(\r,-0.2){{\bf \i}}}
\endpspicture
\vspace{.65cm}

\pspicture(5,1)
\groc{.5,.5}
\qubit{1.5,.5}
\none{2.5,.5}
\none{3.5,.5}
\groc{4.5,.5}
\psset{linewidth=2pt}
\rput(-1,.5){\em b)}
\psline[linestyle=dotted,linecolor=red,dotsep=1pt](-.5,1)(0,1)
\psline[linecolor=red](0,1)(2,1)(2,0)(4,0)(4,1)(5,1)
\psline[linestyle=dotted,linecolor=red,dotsep=1pt](5,1)(5.5,1)

\multido{\r=.5+1,\i=-1+1}{5}{\rput(\r,-0.2){{\bf \i}}}
\endpspicture
\vspace{.65cm}

\pspicture(5,1)
\psset{linewidth=2pt}
\rput(-1,.5){\em c)}
\psline[linestyle=dotted,linecolor=red,dotsep=1pt](-.5,1)(0,1)
\psline[linecolor=red](0,1)(2,1)(5,1)
\psline[linestyle=dotted,linecolor=red,dotsep=1pt](5,1)(5.5,1)
\psset{linewidth=1pt}

\groc{.5,.5}
\qubit{1.5,.5}
\qubit{2.5,.5}
\groc{3.5,.5}
\groc{4.5,.5}
\multido{\r=.5+1,\i=-1+1}{5}{\rput(\r,-0.2){{\bf \i}}}

\endpspicture
\vspace{.65cm}

\end{center}

\caption{ Schematic figure of the measurement protocol for the 2 qubit gate in 1D arrays. {\em a)} The qubits are placed in next-to-neighbour sites. {\em b)} One qubit is left to evolve freely on a size-2 block, while the other is fixed at the boundary. The interaction between the free and the fixed qubit provides a relative phase between the singlet and the triplet conditional on the outcome when measuring the position of the free qubit. {\em c)} The two qubits coupled by a $SWAP$ interaction allows to implement any desired relative phase. The red line represents the potential barrier for the free qubit (not measured). Colour scheme: Yellow represents a site being measured, yielding outcome $\ket0\bra0$. Blue represents outcome $\openone-\ket0\bra0$. Dashed sites are not measured.}
\label{fig:2qbgate}
\end{figure}

In this section we address the dynamics of two qubits when one of them is placed inside a block (the \emph{free} qubit) and the other qubit is localized in one boundary of the block (the \emph{fixed} qubit), (Fig. \ref{fig:2qbgate}\emph{b}). We will show that in this situation, the dynamics for the position of the free qubit is no longer independent of its qubit state. Instead, the position becomes entangled with the two-qubit state. This, combined with position measurements at appropriate times, provides a rich spectrum of two-qubit operations, from unitary two-qubit gates, to projective two-qubit parity measurements. All of these are achieved by position measurements of the qubit inside the block, which in turn are implemented by local projective measurements of the form of Eq.~\eref{eq:kmeasurement}.

Let us develop in more detail some general features of the dynamics for the situation mentioned above. Consider a system containing two qubits ($\mathbb C^4$), located at sites $0$ and $k$ (Fig.~\ref{fig:2qbgate}\emph{a}), where site $0$ is being constantly measured, obviously with outcome $E_\pin^0=\openone-\ket0\bra0$, and the other qubit is left to move freely within the block (Fig.~\ref{fig:2qbgate}\emph{b}). The effective dynamics is given by $H_\eff=\mathcal PH\mathcal P$, where $\mathcal P$ corresponds to the measurement outcome, $\mathcal P=E_\pin^{0}\otimes\openone_{{1,2,\ldots,n}}\otimes E_\pout^{n+1}$, and $H=\sum_{i=0}^nW_{i,i+1}$. Other terms in the Hamiltonian corresponding to sites further away from the block are irrelevant for the dynamics of the block under consideration and only contribute irrelevant global phase factors. Notice that, in analogy with the notation used in the previous sections, the Hilbert space in this situation can be decomposed as a two-qubit space $\mathcal Q^{\otimes 2}\equiv\mathbb C^4$ and $\mathcal L\equiv \mathbb C^n$ for the position of the free qubit. States will be denoted $\ket\Psi=\ket\psi\ket{{\bf k}}$ where $\ket\psi\in\mathcal Q^{\otimes 2}$ and $\ket{{\bf k}}\in\mathcal L$. As in previous sections, $\{\ket{\bf k}\}$~represents an eigenbasis for a position measurement $\pos$, of the free qubit. Using $\mathcal W$ for the two-qubit $SWAP$ gate, one can check that
\begin{eqnarray}
\nonumber
	H_\eff\ket\psi\ket{{\bf 1}}&=&J\left[(\mathcal W+\openone)\ket\psi\ket{\bf 1}+\ket\psi\ket{\bf 2}\right],\\
\nonumber
	H_\eff\ket\psi\ket{{\bf k}}&=&J\ket\psi(\ket{\bf{k-1}}+\ket{\bf{k+1}})\quad 1<k<n,~~~\\
	H_\eff\ket\psi\ket{{\bf n}}&=&J\ket\psi\ket{\bf{n-1}},
\end{eqnarray}
which can be rewritten as
\begin{equation}
	\label{eq:effHam2qbg}
	H_\eff=S\otimes \hat G_++A\otimes \hat G_-,
\end{equation}	
where $S$ ($A$) are projectors onto the symmetric (antisymmetric) subspaces of $\mathcal Q^{\otimes 2}$, respectively, and $\hat G_\pm$ are the conditional position Hamiltonians acting on $\mathcal L$. The specific form for $\hat G_\pm$ is irrelevant for now. The time evolution operator is
\begin{equation}
	\label{eq:2qbdynamics}
	\mathcal U(t)=S\otimes \hat U_+(t)+A\otimes \hat U_-(t),
\end{equation}
where $\hat U_\pm(t)=e^{-i t\hat G_\pm}$ describes the conditional dynamics for eigenstates of the $S$ projector. In general, the symmetric and antisymmetric components will undergo different evolution and a general two-qubit state will become entangled with the position degree of freedom, as suggested by Eq.~\eref{eq:2qbdynamics}.

\begin{figure}
\begin{center}
\includegraphics[width=\linewidth]{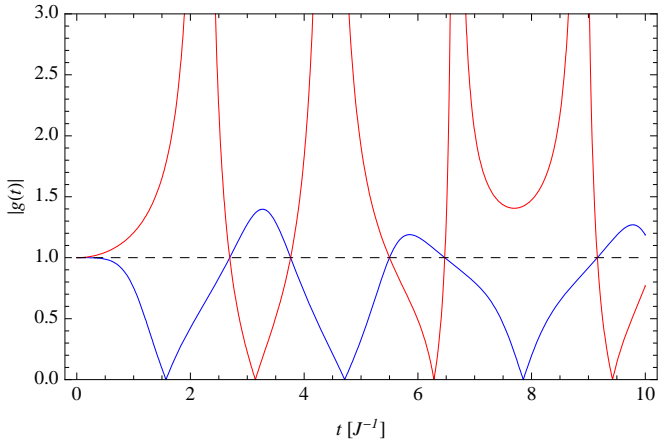}
\caption{The functions $|g_{11}(t)|$ and $|g_{22}(t)|$ (blue) and $|g_{21}(t)|$ and $|g_{12}(t)|$ (red) as a function of time in units of $J^{-1}$. Turning on and off the interaction ensuring unitary two-qubit operations can be achieved when these functions are 1 (dashed line).}
\label{fig:gPlot}
\end{center}
\end{figure}

Let us define $f^\pm_{rk}(t)=\bra{{\bf r}}\hat U_\pm(t)\ket{{\bf k}}$ and $g_{rk}(t)=|g_{rk}(t)|e^{i\varphi_{rk}(t)}=f^-_{rk}(t)/f^+_{rk}(t)$ [see Fig.~\ref{fig:gPlot}]. The functions $f^\pm_{rk}(t)$ are the transition amplitudes for the free qubit being in position ${r}$ after time $t$, when the initial position was ${k}$, conditional on the parity of the two-qubit state.

\begin{figure}
\begin{center}
\includegraphics[width=\linewidth]{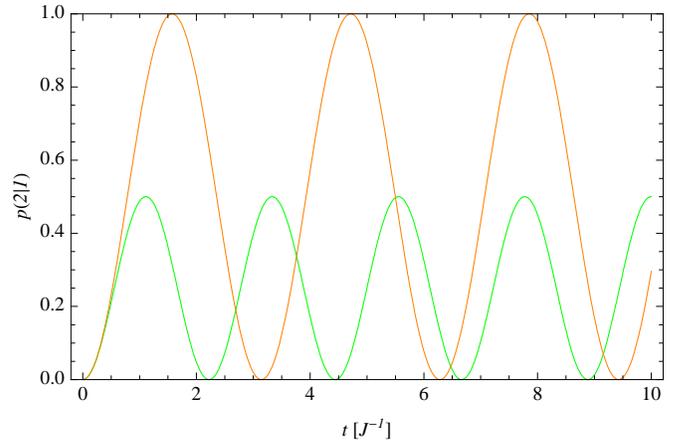}
\caption{Success probability $p(1|2)=p(2|1)$ for switching on/off the interaction when the two qubits are in the triplet state (green) and in the singlet state (orange). The probability for the singlet equals that of the triplet whenever $|g_{11}(t)|=1$. At these times, the measurement does not provide any information about the parity of the two-qubit state.}
\label{fig:probPlot}
\end{center}
\end{figure}

Performing position measurements $\pos$ transforms the two-qubit state whenever the position is entangled with the two-qubit state. A position measurement for the free qubit at time $t$, when the initial state was $\ket\Psi=\ket\psi\ket{\bf k}$ yields outcome $r$ with probability
\begin{eqnarray}
\nonumber
	p({\bf r}|{\bf k})&=&\left\|\bra{{\bf r}}\mathcal U(t)\ket\psi\ket{\bf k}\right\|^2\\
\nonumber
	&=&\left\| \left(f^+_{rk}(t)S+f^-_{rk}(t)A\right)\ket\psi\right\|^2\\
	&=&|f^+_{rk}(t)|^2\langle S\rangle+|f^-_{rk}(t)|^2\langle A\rangle,
\end{eqnarray}
where $\langle S\rangle=\bra\psi S\ket\psi$ ($\langle A\rangle=\bra
\psi A\ket\psi$) is the symmetric (antisymmetric) component of $\ket\psi$ [see Fig.~\ref{fig:probPlot}]. The two-qubit state after the measurement is
\begin{equation}
	\label{eq:2qbcollapse}
	\ket{\psi'}=\frac{f^+_{rk}(t)}{\sqrt{p({\bf r}|{\bf k})}}\left(S+{g_{rk}(t)A}\right)\ket\psi.
\end{equation}

The effect of the measurement is therefore determined by the initial and final positions, $k$ and $r$ respectively, and the time at which the measurement is performed. In general, a position measurement can indirectly learn about the two-qubit state, with the consequent collapse of the state vector, Eq.~\eref{eq:2qbcollapse}.

\subsection{Two qubit gates}
\label{sec:2qbgates}
A natural question is whether two-qubit gates can be implemented within this scheme. In this section we answer this question in the positive. The nature of our Hamiltonian clearly suggests that we aim at implementing a $\sqrt{\mathcal W}$ gate. More generally, let $\mathcal W(\phi)=S+e^{i\phi}A$ (with $\mathcal W(\pi/2)=\sqrt{\mathcal W}$). If two qubits could be placed in confined neighbouring sites (a size-two block containing two qubits) and separated without further difficulty, implementation of $\mathcal W(\phi)$ for arbitrary $\phi$ would be trivial. One would just let the system evolve for a time $t=\phi/J$, and the quantum gate $\mathcal W(\phi)$ would be implemented. However, placing the two qubits in neighbouring sites and separating them without affecting the two-qubit state is not possible. Despite this difficulty we show in the following a procedure to implement any two-qubit gate of the form $\mathcal W(\phi)$, with arbitrary $\phi$.

Let us consider the starting configuration as depicted in Fig. \ref{fig:2qbgate}\emph{a}. This is a static configuration that can be easily reached with the state transfer procedures explained above. The two-qubit gate will be implemented through three stages;  $1)$ \emph{interaction switch on}, $2)$ \emph{interacting}, and $3)$ \emph{interaction switch off}. The \emph{switch on/off} stages use a repeat-until-success method, while the \emph{interacting} stage is deterministic and is used to accommodate arbitrary $\phi$ phases and correct the phases introduced by the \emph{switch on/off} stages. Even the phases introduced in the unsuccessful switch-off attempts can be corrected by remaining in the interacting stage a given time after each failed attempt.

The switch on/off stages consist on letting the free qubit propagate and performing a position measurement at specific times. It is clear that in order to preserve coherence in the switch on/off stages one must ensure that for every possible outcome $r$ of the position measurement, the transformation on the two-qubit state be unitary, \emph{i.e.,}
\begin{equation}\footnotesize
	\left[\frac{f^+_{rk}(t)}{\sqrt{p({\bf r}|{\bf k})}}\left(S+{g_{rk}(t)A}\right)\right]^\dagger=\left[\frac{f^+_{rk}(t)}{\sqrt{p({\bf r}|{\bf k})}}\left(S+{g_{rk}(t)A}\right)\right]^{-1},\quad \forall \bf{r}.
\end{equation}
A little algebra shows that this requirement is equivalent to
\begin{equation}
	\label{eq:unitarycondition}
	|g_{rk}(t)|=1\qquad\forall \bf r,
\end{equation}
which implies that $|f^+_{rk}(t)|=|f^-_{rk}(t)|$ for all $r$ [see Fig.~\ref{fig:gPlot}]. This means that $p(\bf r|\bf k)$ must be independent of the parity of the two-qubit state. Moreover, limiting the size of the block to two sites guarantees that $|g_{1k}(t)|=1 \Longleftrightarrow |g_{2k}(t)|=1$, as follows from
\begin{equation}
	|g_{1k}(t)|^2=\frac{|f^-_{1k}(t)|^2}{|f^+_{1k}(t)|^2}=\frac{1-|g_{2k}(t)|^2|f^+_{2k}(t)|^2}{1-|f^+_{2k}(t)|^2}.
\end{equation}
Let $t^*$ be such that $|g_{1k}(t^*)|=|g_{2k}(t^*)|=1$. Performing a position measurement on the free qubit at time $t^*$ does not provide any information about the two-qubit state. However, the state is transformed according to $\mathcal W(\varphi_{rk}(t^*))\ket\psi$, where $k$ is the initial position of the free qubit and $r$ is the outcome of the measurement.

At this point, it is clear that the switch on stage introduces a relative phase $m\varphi_{22}(t^*)+\varphi_{12}(t^*)$, where $m$ is the number of failed attempts. This phase can easily be corrected during the interacting stage, when also the target phase $\phi$ is applied. The switch off stage will introduce, when successful, a phase $\varphi_{21}(t^*)$, that can be also accounted for during the interacting stage. Therefore, in the interacting stage the total phase that needs to be applied is $\phi-m\varphi_{22}(t^*)-\varphi_{12}(t^*)-\varphi_{21}(t^*)$. When the switch off stage is unsuccessful, the introduced phase is $\varphi_{11}(t^*)$ and the qubits remain coupled. Hence, the phase can be corrected immediately after each unsuccessful switch off stage. The resulting two-qubit state after successful switch off is
\begin{equation}
	\ket{\psi'}=\mathcal W(\phi)\ket\psi.
\end{equation}
See a sketch of the algorithm in Fig.~\ref{fig:2qbg_alg}.

\begin{figure}\flushleft{
\verb|// TWO-QUBIT GATE ALGORITHM //|\\
\verb|PLACE qubits in next-to-nearest sites.| (Fig.~\ref{fig:2qbgate}\emph{a})\\
\verb|SET |$\theta:=\phi$\\ 
\verb|DO {|\\
\verb|    LET one qubit propagate freely between site 2|\\
\verb|        and site 1 for a time |$t^*$ (Fig.~\ref{fig:2qbgate}\emph{b})\\
\verb|    MEASURE qubit position (|$\mapsto$\verb|k)|\\
\verb|    SET |$\theta:=\theta-\varphi_{k2}(t^*)$\\
\verb|} UNTIL k=1|\\
\verb|DO {|\\
\verb|    LET the two qubits interacting for a time|\\
\verb|        |$t=J^{-1}(\theta-\varphi_{21}(t^*)~\textrm{mod}2\pi)$~~~~(Fig.~\ref{fig:2qbgate}\emph{c})\\
\verb|    LET one qubit propagate freely between site 1|\\
\verb|        and site 2 for a time |$t^*$ (Fig.~\ref{fig:2qbgate}\emph{b})\\
\verb|    MEASURE qubit position (|$\mapsto$\verb|k)|\\
\verb|    SET |$\theta:=\varphi_{21}(t^*)-\varphi_{k1}(t^*)$\\
\verb|} UNTIL k=2|\\
\verb|END|
}
\vspace{10pt}
\caption{Pseudocode algorithm for the two-qubit gate $\mathcal {W}(\phi)$. Variable $\phi$ is the target relative phase, $\theta$ is the phase remaining to be applied. This is updated whenever some relative phase is applied to the two-qubit state. The algorithm ends when the free qubit is found in site {2} after having been in site {1}. On exit, the remaining phase is $\theta=0$. $t^*$ must be chosen such that $|g_{rk}(t^*)|=1$.}
\label{fig:2qbg_alg}
\end{figure}

Notice that this scheme is deterministic, \emph{i.e.,} it can be implemented with a finite number of trials with exponentially close to 1 probability. Furthermore, the results developed so far hold with full generality with arbitrary setups where size-2 blocks can be constructed. This means that other geometries such as 2D lattices, ladders or zigzag chains can also be approached with the same analysis. For the particular case at hand we have
\begin{eqnarray}
	\hat G_+&=& \hat{\bf{X}}+2\ket{{\bf{1}}}\bra{{\bf{1}}},\\
	\hat G_-&=& \hat{\bf{X}},
\end{eqnarray}
and
\begin{eqnarray}
|g_{11}(t)|=|g_{22}(t)|&=&\frac{2\left|\cos t\right|}{\sqrt{3+\cos2\sqrt2t}},\\
|g_{11}(t)|=|g_{22}(t)|&=&\sqrt2\left|\csc \sqrt2 t~\sin t\right|.
\end{eqnarray}

Moreover, when computing the time $t^*$ for the target phase, there are several solutions. Thus, one can choose several times at which the target phase is acquired, while optimizing for success probability, number of trials or other considerations. 

Let us remark that by implementing $\mathcal W(\pi)=\mathcal W$ gates, one can interchange the positions of two qubits, thus allowing for arbitrary two-qubit crossings. This allows us to perform perfect state transfer not only inside a block but in the whole register. 

\subsection{Parity measurements}
\label{sec:2qbparity}
Just like in the previous section we have shown how position measurements can induce unitary transformations on the two-qubit state by choosing to perform the measurements at appropriate times, projections on the symmetric and antisymmetric two-qubit subspaces (triplet and singlet, respectively) can equally be implemented by measuring at times $t^*$ when the effect of the measurement is an $S$ or $A$ projection. In particular we show that choosing $t^*$ appropriately one can implement a generalized measurement with one conclusive outcome ($S$ or $A$) and one inconclusive outcome. By alternating these two kinds of measurement, one obtains, with exponentially close to 1 probability in the number of trials, a conclusive outcome corresponding to $S$ or $A$.

Let the system be prepared in the configuration of Fig.~\ref{fig:2qbgate}\emph{a}. Choose $t_1$ such that $f^-_{12}(t_1)=0$. By performing a position measurement at time $t_1$ the effect of the measurement on the two-qubit state can be described by $\{K_1^{(1)}, K_2^{(1)}\}$ where
\begin{eqnarray}
	K_1^{(1)}&=&f^+_{12}(t_1)S,\\
	K_2^{(1)}&=&f^+_{22}(t_1)S+f^-_{22}(t_1)A,
\end{eqnarray}
where subindices indicate the outcome (site where the free qubit is encountered) and superscript $(1)$ indicates that the measurement is performed after time $t_1$ of free evolution. An outcome in site 1 corresponds to a projection onto the symmetric subspace $S$, while if the qubit is encountered in site $2$, no conclusive outcome can be inferred. However, one can perform another measurement after time $t_2$ such that $f^+_{12}(t_2)=0$, with corresponding Kraus operators
\begin{eqnarray}
	K_1^{(2)}&=&f^-_{12}(t_2)A,\\
	K_2^{(2)}&=&f^+_{22}(t_2)S+f^-_{22}(t_2)A.	
\end{eqnarray}
With this, the overall final effect on the two-qubit state can be summarized as
\begin{eqnarray}
	K_S&=&K_1^{(1)}=f^+_{12}(t_1)S,\\
	K_A&=&K_1^{(2)}K_2^{(1)}=e^{i\phi_1}f^-_{12}(t_2)A,\\
	K_\star&=&K_2^{(2)}K_2^{(1)}=e^{i\phi_2}f^+_{22}(t_1)S+e^{i\phi_1}f^-_{22}(t_2)A,
\end{eqnarray}
where $\star$ is used to denote an inconclusive outcome, and $e^{i\phi_1}\equiv f^-_{22}(t_1)$ and $e^{i\phi_2}\equiv f^+_{22}(t_2)$, which are guaranteed to be pure phases due to the choice of $t_1$ and $t_2$ and unitarity of $U_\pm(t)$. The corresponding POVM operators are $E_x=K_x^\dagger K_x$, fulfilling $\sum_x E_x=\openone$ and yielding the probabilities $p(x)=\bra\psi E_x\ket\psi$,
\begin{eqnarray}
	p(S)&=&|f^+_{12}(t_1)|^2\langle S\rangle,\\
	p(A)&=&|f^-_{12}(t_2)|^2\langle A\rangle,\\
\nonumber
	p(\star)&=&1-p(S)-p(A)\\
	&=&|f^+_{22}(t_1)|^2\langle S\rangle+|f^-_{22}(t_2)|^2\langle A\rangle.
\end{eqnarray}
The free qubit remains in site $2$ after every inconclusive outcome, while hopping to site 1 indicates a conclusive outcome depending on the stage of the protocol in which the hopping occurs ($S$ or $A$ for hopping in the first or second measurement, respectively). This measurement scheme can be iterated until a conclusive outcome is obtained. After a conclusive outcome it is trivial to separate the two qubits by just enlarging the block size and letting the free qubit propagate for some time.  

It is important to notice that after every inconclusive outcome the two-qubit state is altered as $\ket\psi\rightarrow K_\star\ket\psi$. The Kraus operators corresponding to outcome $S$ or $A$ after $n$ inconclusive outcomes are given by
\begin{eqnarray}
	K_{S,n}&=&K_SK_\star^n=e^{in\phi_2}f_{12}^+(t_1)(f_{22}^+(t_1))^nS\\
	K_{A,n}&=&K_AK_\star^n=e^{i(n+1)\phi_1}f_{12}^-(t_2)(f_{22}^-(t_2))^nA
\end{eqnarray}
which, as required, fulfill $\sum_{x,n}K_{x,n}^\dagger K_{x,n}=\openone$.

\begin{figure}[t]\flushleft{
\verb|// TWO-QUBIT PARITY MEASUREMENT ALGORITHM //|\\
\verb|PLACE qubits in next-to-nearest sites.| (Fig.~\ref{fig:2qbgate}\emph{a})\\
\verb|DO {|\\
\verb|    LET qubit propagate freely between site 2|\\
\verb|        and site 1 for a time |$t_1$ (Fig.~\ref{fig:2qbgate}\emph{b})\\
\verb|    MEASURE qubit position (|$\mapsto$\verb|k)|\\
\verb|    IF k=1 THEN|\\
\verb|        outcome = S|\\
\verb|    ELSE|\\
\verb|        LET qubit propagate freely between site 2|\\
\verb|            and site 1 for a time |$t_2$ (Fig.~\ref{fig:2qbgate}\emph{b})\\
\verb|        MEASURE qubit position (|$\mapsto$\verb|k)|\\
\verb|        IF k=1 THEN|\\
\verb|            outcome = A|\\
\verb|        ELSE|\\
\verb|            outcome = |$\star$\\
\verb|} UNTIL outcome|$\,\neq\star$\\
\verb|END|
}
\vspace{10pt}
\caption{Algorithm for implementing the two-qubit parity measurement written in pseudocode. The procedure is repeated until a conclusive outcome is obtained. The loop consists on alternating weak $S$ and $A$ projections. After the conclusive outcome the two-qubit state is collapsed onto the singlet/triplet subspace and the free dynamics only provides irrelevant global phases. Hence the two qubits can be decoupled easily with the same tools.}
\label{fig:parity_alg}
\end{figure}

It is relevant to consider the probability for a given number of consecutive inconclusive outcomes. The probability for $n$ successive inconclusive outcomes is given by 
\begin{equation}
	p(\star^n)=\|K_\star^n\ket\psi\|^2=|f^+_{22}(t_1)|^{2n} \langle S\rangle+|f^-_{22}(t_2)|^{2n}\langle A\rangle, 
\end{equation}
which decreases exponentially in $n$. More importantly, the probability of obtaining $n$ inconclusive outcomes followed by a conclusive one is
\begin{eqnarray}
	p(S\star^n)&=&|f^+_{12}(t_1)|^2|f^+_{22}(t_1)|^{2n}\langle S\rangle,\\
	p(A\star^n)&=&|f^-_{12}(t_2)|^2|f^-_{22}(t_2)|^{2n}\langle A\rangle.
\end{eqnarray}
Summing over all possible number of inconclusive outcomes yields
\begin{eqnarray}
	P(S)&=&\sum_{n=0}^\infty p(S\star^n)=\frac{|f^+_{12}(t_1)|^2}{1-|f^+_{22}(t_1)|^{2}}\langle S\rangle=\langle S\rangle,
\end{eqnarray}
and analogously $P(A)=\langle A\rangle$, as expected from a quantum mechanical parity measurement. More formally, it can be seen that the superoperator associated with outcome $x$ ($=S$ or $A$), given by $\mathcal K_x\rho=\sum_n K_{x,n}^\dagger \rho K_{x,n}$ is
\begin{eqnarray}
	\mathcal K_S\rho=S\rho S,\qquad\qquad\mathcal K_A\rho=A\rho A,
\end{eqnarray}
which is nothing but a parity measurement. Moreover, since parity is a conserved quantity in the effective dynamics, this sequence of measurements implements a QND parity measurement~\cite{braginsky}.

Therefore, a projective parity measurement that distinguishes the singlet state from the triplet can be implemented by a trial and error generalized measurement implemented through local measurements of the form of~Eq.~\eref{eq:kmeasurement}. See the pseudocode of the algorithm in Fig.~\ref{fig:parity_alg}

As in the two-qubit gate, our analysis is completely general and applies to other lattice geometries. Also, the choice for timings corresponds to arbitrary choices of assigning position outcomes to symmetric or antisymmetric outcomes. Different choices could be made which take into account other considerations such as expected number of trials before a conclusive outcome, average time before a conclusive outcome occurs, robustness against fluctuations in some parameter, etc. The aim of this section is not to engineer a specific protocol but rather to show that, in principle, the parity measurement is possible within a finite time given the assumed Hamiltonian dynamics and the projective measurements of~Eq.~\eref{eq:kmeasurement}.

\section{Extensions to the Hamiltonian: creating and rotating qubits}
\label{sec:extensions}

In previous sections we have seen that, in principle, perfect state transfer, two-qubit gates and projective parity measurements can be implemented only by making use of assumptions \emph{a)} and \emph{b)} in Section~\ref{sec:intro}. 
Although these are certainly crucial tasks, these assumptions alone do not provide universal quantum computation. Hence, it would be useful to know in what directions must these assumptions be slightly extended, should one be interested in achieving universal quantum computation. To that effect, it is relevant to investigate possible extensions to  \emph{a)} and \emph{b)}, which provide additional tasks; in particular we study how to obtain  \emph{i)} spontaneous maximally entangled pair creation which can be used to initialize the quantum register, and \emph{ii)} single qubit rotations.

\subsection{Spontaneous maximally entangled qubit pair creation}
So far we have concentrated in the Hamiltonian of Eq.~\eref{eq:GenHeisHam} where $\theta=\pi/4$. A reasonable question is whether detuning the parameter $\theta$ provides any useful dynamics. The Hamiltonian then becomes
\begin{eqnarray}
\nonumber
	H&=&\tilde J\sum_i \left(\cos \theta~\vec S_i \vec S_{i+1}+\sin \theta (\vec S_i \vec S_{i+1})^2\right)\\ 
\nonumber
	&=&\tilde J \sin\theta ~\openone+\tilde J\cos\theta\sum_i W_{i\,i+1}\\
\label{eq:GenHeisHam2}
	&&-3\sqrt2\tilde J\sin\left(\frac\pi4-\theta\right)\sum_i\ket\chi_{i\,i+1}\bra\chi,
\end{eqnarray}
where we have used the identity
\begin{eqnarray}
	\vec S_i\vec S_{i+1}&=&W_{i\,i+1}-3\ket\chi_{i\,i+1}\bra\chi,\\
\label{eq:bilbiq}
	\vec S_i\vec S_{i+1}+(\vec S_i\vec S_{i+1})^2&=&W_{i\,i+1}+\openone,
\end{eqnarray}
and we have defined
\begin{equation}
	\ket\chi=\frac{\ket{\up\downarrow}+\ket{\down\uparrow}-\ket{00}}{\sqrt3}.
\end{equation}
Eq.~\eref{eq:bilbiq} contributes to an irrelevant global phase factor ($\tilde J\sin\theta~\openone$) and could have been omitted from Eq.~\eref{eq:GenHeisHam2}. Eqs.~\eref{eq:GenHeisHam2} and \eref{eq:bilbiq} show that for $\theta$ values different from $\pi/4$, entangled qubit pairs in the state $\ket{\psi^+}={1\over\sqrt{2}}\left(\ket{\up\downarrow}+\ket{\down\uparrow}\right)$ can be spontaneously created.

Concentrating on the effective dynamics for size-2 blocks, one can easily see that
\begin{eqnarray}
&&\mathcal P\left(\sum_{i=0}^2\ket\chi_{i\,i+1}\bra\chi\right)\mathcal P=\\
\nonumber
&&\ket0_{0}\bra0\otimes\Big(\ket{0}\bra{0}\otimes\openone+\ket\chi\bra\chi+\openone\otimes\ket{0}\bra{0}\Big)_{12}\otimes\ket0_3\bra0,
\end{eqnarray}
with $\mathcal P=\ket0_{0}\bra0\otimes\openone_1\otimes\openone_2\otimes\ket0_3\bra0$. With this, one can see that the dynamics for single qubits in size-2 blocks remains unchanged. The term $\ket{0}\bra{0}\otimes\openone+\openone\otimes\ket{0}\bra{0}$ corresponds to the identity $\openone\otimes\openone$ operator in the $\mathcal Q\otimes\mathcal L$ representation, hence contributing only to a global phase. The term $\ket\chi\bra\chi$ trivially vanishes in the one-qubit subspace.

However, for empty two-site blocks, the dynamics becomes nontrivial, as the term $\ket\chi\bra\chi$ couples the states $\ket{00}$ to the state $\ket{\psi^+}$. More precisely, the transition probability from an empty block to an occupied block is\begin{equation}
	\left|\bra{\psi^+}\exp\left(-it H\right)\ket{00}\right|^2=\frac{8}{9}\sin^2\left(\frac{3t\lambda}{2}\right),
\end{equation}
where $\lambda=3\sqrt2\tilde J\sin\left(\pi/4-\theta\right)$. Hence, measuring the number of particles after a time $t=\pi/(3\lambda)$ one will obtain two qubits in the entangled state $\ket{\psi^+}$ with probability $8/9$.

This shows that, given an initialization of the spin chain to $\ket0^{\otimes n}$, tweaking the parameter $\theta$ in the Hamiltonian, combined with continuous projective measurements of the form of Eq.~\eref{eq:kmeasurement} provide a means to create pairs of qubits from the initially \emph{empty} chain. Notice that this is a global modification of the total Hamiltonian, which does not imply a site-wise modification of the interaction. Moreover, this modification only needs to be made at a beginning stage of the protocol, when qubits are created for later information storage/transfer/processing. Moreover, the procedure directly creates entangled pairs that can be deterministically separated using the methods established in Section~\ref{sec:2qbdynamics}, and distributed by means of the state transfer techniques and therefore used as resources for teleportation.

\subsection{Single qubit operations}
\label{sec:sqgates}
Single qubit gates are, in principle, an easy task, since they do not require many-body interactions. If one can assume local addressability, it is straightforward to devise schemes in which the qubit is localized by the $\pos$ measurement and the corresponding external field is applied. This assumes that the two qubit states $\{\ket\up,\ket\down\}$ can be selected. However, assuming local measurements of the form of Eq.~\eref{eq:kmeasurement} does not imply that all sorts of local addressability are granted. For this reason, we would like to address the situation in which the only local addressing is done via the measurements of Eq.~\eref{eq:kmeasurement} and any other perturbation is as general as possible.

With full generality, any external perturbation that locally transforms the states in an $n$ site block can be expressed as an additional term in the Hamiltonian of the form
\begin{equation}
	\label{eq:extinteraction}
	H_{\textrm{ext}}=\sum_{k=1}^n \sum_{\mu=1}^8c_k^\mu \lambda_k^\mu,
\end{equation}
where $\lambda_k^\mu$ is the $\mu$th Gell-Mann matrix acting on site $k$. For simplicity, let us assume that the generators are expressed in the basis $\{\ket\up,\ket\down,\ket 0\}$ (in this order). Assuming that the block contains a single qubit, upon continuous measurement of the $\pos$ observable with outcome $k$, the effective Hamiltonian becomes
\begin{equation}
	  H'_{\eff }=\sum_{\mu=1}^3 c^\mu_k   \sigma^\mu_k,
\end{equation}
where $k$ is the outcome of the $\pos$ measurement. This reduction is easily seen by observing that $\mathcal P \lambda_r^\mu\mathcal P$ is trivial for all sites $r\neq k$ ($\ket0\bra0\lambda^\mu\ket0\bra0=0$ for $1\leq\mu\leq7$ and $\ket0\bra0\lambda^8\ket0\bra0=-{2\over\sqrt3}\ket0\bra0$), whereas the terms corresponding to the outcome of $\pos$, $\mathcal P\lambda_k^\mu\mathcal P$ are
\begin{eqnarray}
	(\openone-\ket0\bra0)\lambda_k^i(\openone-\ket0\bra0)&=&\sigma_i\qquad i=1,2,3\\
	(\openone-\ket0\bra0)\lambda_k^i(\openone-\ket0\bra0)&=&0\qquad i=4,5,6,7\\
	(\openone-\ket0\bra0)\lambda_k^8(\openone-\ket0\bra0)&=&\frac{1}{\sqrt3}\openone.
\end{eqnarray}
This means that only the coefficients $c_k^\mu$ for $k$ being the outcome of the $\pos$ measurement and $\mu=1,2,3$ are relevant for the evolution of the qubit, which is generated by the effective Hamiltonian
\begin{equation}
	H_\eff=\vec c\cdot\vec \sigma\otimes\openone
\end{equation}
in the $\mathcal Q\otimes\mathcal L$ representation.

The initial Hamiltonian assumed here is relatively general and will accommodate several practical situations. In particular, it is remarkable that local access to single sites is only required for the measurements, whereas the external Hamiltonian does not require to be of single-site or local nature. Although it is not our aim here to suggest a physical implementation for our scheme, it is worth stressing that this level of generality accommodates the reality of several physical implementations. Moreover, the scheme naturally prevents transitions between the $\ket\up$, $\ket\down$ states and the $\ket0$ state.

As an illustrative example we can show how to describe a site-dependent magnetic field $\vec B_k$ coupled to the spin-1 chain. The interaction Hamiltonian reads 
\begin{equation}
	\sum_k \vec B_k\cdot \vec S_k.
\end{equation}
The spin-1 matrices can be written as
\begin{equation}
	S_x=\frac{1}{\sqrt2}(\lambda^4+\lambda^6),\quad S_y=\frac{1}{\sqrt2}(\lambda^5-\lambda^7),\quad S_z=\lambda^3.
\end{equation}
Notice that the spin matrices need to be properly rearranged to meet the basis ordering that we have picked $\{\ket\up,\ket\down,\ket0\}$. Once we have this, we can cast the magnetic interaction into the form of Eq.~\eref{eq:extinteraction} by defining
\begin{eqnarray}
	c^1_k=c^2_k=c^8_k=0,\qquad
	&c^3_k=B_k^z,\\
	c^4_k=c^6_k=\frac{B_k^x}{\sqrt2},\qquad
	&c^5_k=-c^7_k=\frac{B_k^y}{\sqrt2},	
\end{eqnarray}
which yields an effective Hamiltonian
\begin{equation}
	H_\eff=B_k^z~\bf Z\otimes\openone.
\end{equation}
This also shows that a homogeneous magnetic field introduces a constant relative phase between the $\ket\up$ and $\ket\down$ states, which can be accounted for by going to the interaction picture.

Note that the particular form of the interaction depends on the choice of the \emph{vacuum} state $\ket0$. A completely analogous derivation shows that, if the \emph{vacuum} representative had been chosen to be $\ket{\down}$ instead of $\ket0$, the effective Hamiltonian would be
\begin{equation}
	H_\eff=\left(\frac{B^x_k}{\sqrt2}{\bf X}+\frac{B^y_k}{\sqrt2}{\bf Y}+\frac{B^z_k}{2}{\bf Z}\right)\otimes\openone.
\end{equation}
This choice, however, would also affect the results of the previous subsection, and entangled qubit pairs would no longer be easily created. Nevertheless, the results on state transfer and two-qubit gates do remain unchanged.

%

\section{Concluding remarks} 
\label{sec:conclusions}

We have shown how one can implement, in a one-dimensional quantum many-body system, a quantum register in which quantum state transfer, universal two-qubit gates, and two-qubit parity measurements can be achieved by an always-on, and time-independent, many-body Hamiltonian with nearest-neighbour interactions. The control required for performing these tasks, which are the crucial ingredients for universal quantum computation, is provided by frequent projective measurements (yielding the quantum Zeno effect) at the appropriate times and sites of the chain. Moreover, measurements are always in the same basis.
We have also shown some possible modifications of the initially assumed dynamics in order to create maximally entangled qubit pairs, which can be used to initialize the quantum register, and single qubit gates. 

The ideas presented here can be easily generalized to 2 and 3-dimensional lattices. Although 3D lattices would provide the best scalability, we believe 2D would provide the optimal trade-off between scalability and difficulty to implement the measurements. Moreover, 2D lattices have the advantage over 1D that the qubits can be freely moved without the need to perform a $SWAP$ gate at every qubit crossing. Also, different configurations such as ladders or zigzag chains may provide interesting geometries.

The natural and interesting extension of this work is to present an experimental proposal of the theoretical scheme presented here. This would open up many interesting questions such as the robustness of the scheme and the effect of imperfect measurements. To this purpose, we envisage two suitable physical systems: 1) Coupled arrays of quantum dots~\cite{quantumdots}, which naturally implement the $t$-$J$ model, of which our Hamiltonian is a particular case ($t=J$). The charge electrons would play the role of the qubit, each site having three possible states, one in which the dot is empty and two for the occupied dot with spin-up and spin-down electron. The on-site measurements would be performed by charge-sensing techniques such as quantum point contacts. On the other hand the exchange coupling and the hopping rates could be tuned independently, giving extra freedom to implement two-qubit gates without the need to resort to repeat-until-success methods. 2) Trapped ions, where the Hamiltonian could be implemented by effective methods~\cite{cirac_porras} and the on-site measurements by resonance fluorescence measurements. In this case, one should investigate how to avoid the heating of the ions by photon scattering in the measurement process.

In a general framework, we would like to briefly discuss some error sources hindering the effectivity of our scheme, in particular for the implementation of two-qubit operations. It is difficult to establish a specific error model without specializing to a particular physical implementation. However, three kinds of errors related to imperfect local measurements may be anticipated:
\begin{enumerate}
    \item \emph{Measurements projecting on a basis different than expected}. Since the Hamiltonian is basis-independent this error is prevented, as one may choose the vacuum state as the one distinguishable by the measurement.
    \item \emph{False negatives}, where a measurement fails to yield outcome ``$\pout$'' with small probability.  This error can be reduced by concatenating several measurements, exponentially reducing the probability of a false negative at the expense of increasing the time it takes for a measurement to be performed.
    \item \emph{Uncertainties on the time $t$ at which the measurements are performed}, as a result of technical difficulties or of fundamental limitations imposed by the time-energy uncertainty relations. This source of error is intimately related to the previous one and has the potential effect of removing us from the Zeno regime. A direct way of coping with this error is to reduce the coupling strength $J$ in such a way as to ensure that the Zeno dynamics remain a good approximation. Most measurements are however, continuous in nature (\emph{i.e.}, resonance fluorescence in ions or cold atoms, charge measurements on quantum dots, etc.). Despite the departure from the theoretical framework in which standard von Neuman measurements are often thought of, a continuous time analysis of the quantum Zeno effect can be made, leading to essentially the same conclusions~\cite{facchi_quantum_2007}. The question is then, whether or not one can turn on and off the measurement at a timescale smaller than that of the Hamiltonian dynamics. If that is not the case, a more detailed analysis should be made, which involves the optimization of the \emph{average} times $t^*$, $t_1$ and $t_2$ at which the measurements are performed. Expressing the average gate fidelities as a function of these times [averaged over all possible two-qubit states, number of trials and time fluctuations $\Delta t$], and assuming that fluctuations are small, one can see that the essential quantities contributing to a non-perfect fidelity are the derivatives of the functions $f^\pm(t)$ and $g(t)$. One may exploit the freedom available in choosing the times $t^*$, $t_1$ and $t_2$, in order to reduce the noise in the gates, by finding a compromise between the average or expected number of trials (each trial introduces some error) before the gate is completed, the time derivatives of the relevant functions (quickly varying functions lead to larger errors), and the kinds of errors that one may be able to tolerate (\emph{i.e.}, phase errors, singlet-triplet collapse for controlled phase gates, etc.).
\end{enumerate}
Finally, there are some kinds of errors that cannot be corrected within the proposed setup. The most prominent of them is revealing information regarding the actual state of the qubit (\emph{i.e.}, revealing whether a site is in state $\ket\up$ or $\ket\down$). Such errors cannot be addressed by methods specific to our scheme, and error-correction schemes would then be necessary. Also, non-projective measurements different from the \emph{false negative} or \emph{false positive} have no clear solution within this scheme. However, these seem to be highly unlikely for the kinds of implementations that we have mentioned. In conclusion, the difficulties and possibilities for correcting and tolerating errors should not lead to resignation. Instead, they must serve as a guide to choose the most appropriate physical implementation.\\*

\noindent{\bf Acknowledgements}
 We are thankful to A.~Ac\'in, A.~Beige, S.~Bose, D.~Porras and A.~Sanpera for fruitful discussions. We thank the hospitality of the 2007 QI Workshop in the Benasque Center for Science. This work was supported by spanish MEC grants AP2005-0595, FIS2005-03169, Consolider-Ingenio2010 CSD2006-00019 QOIT, catalan grant SGR-00185, EU IP program SCALA. A.~M. acknowledges financial support from the UK program QIPRC and EC under the FP7 STREP 
Project HIP, Grant Agreement n. 221889.\\*
\noindent


\begin{thebibliography}{000}
\expandafter\ifx\csname natexlab\endcsname\relax\def\natexlab#1{#1}\fi
\expandafter\ifx\csname bibnamefont\endcsname\relax
  \def\bibnamefont#1{#1}\fi
\expandafter\ifx\csname bibfnamefont\endcsname\relax
  \def\bibfnamefont#1{#1}\fi
\expandafter\ifx\csname citenamefont\endcsname\relax
  \def\citenamefont#1{#1}\fi
\expandafter\ifx\csname url\endcsname\relax
  \def\url#1{\texttt{#1}}\fi
\expandafter\ifx\csname urlprefix\endcsname\relax\def\urlprefix{URL }\fi
\providecommand{\bibinfo}[2]{#2}
\providecommand{\eprint}[2][]{\url{#2}}

\bibitem{amico08}
L.~Amico, R.~Fazio, A.~Osterloh, and V.~Vedral, Rev. Mod. Phys. \textbf{80}, 517 (2008).

\bibitem{raussendorf_one-way_2001}
R.~Raussendorf and H.~J.~Briegel, Phys. Rev. Lett. \textbf{86}, 5188 (2001).

\bibitem{adiabatic}
E.~Farhi, J.~Goldstone, S.~Gutmann and M. Sipser, arXiv:quant-ph/0001106

\bibitem{kitaev}
A.~Kitaev, Annals Phys. \textbf{303}, 2-30 (2003).

\bibitem{Zeno}
R.~G. Winter, Phys. Rev. \textbf{123}, 1503 (1961).

\bibitem{Zenob}
B.~Misra and E.~C.~G. Sudarshan, J. Math. Phys. \textbf{18}, 756 (1977).

\bibitem{ULS}  G.~V.~Uimin, JETP. Lett. \textbf{12}, 225 (1970). 

\bibitem{ULSb}
C.~K.~Lai, J. Math. Phys. \textbf{15}, 1675 (1974). 

\bibitem{ULSc}
B.~Sutherland, Phys. Rev. B \textbf{12}, 3795 (1975).

\bibitem{facchi_quantum_2007}
P.~Facchi and S.~Pascazio, J. Phys. A: Math. Theor. {\bf 41} 493001 (2008) and references therein.

\bibitem{benjamin} S.C.~Benjamin, B.W.~Lovett, J.H.~Reina, Phys. Rev. A, \textbf{70}, 060305(R) (2004).

\bibitem{quantum_walks}
E.~Farhi and S.~Gutmann, Phys. Rev. A \textbf{58}, 915 (1998).

\bibitem{quantum_walksb}
J.~Kempe, Cont. Phys. \textbf{44}, 307 (2003).



\bibitem{bose_overview_2007}
S.~Bose, Contemporary Physics \textbf{48}, 13 (2007).

\bibitem{balachandran_quantum_2000} 
A.~P. Balachandran and S.~M. Roy, Phys. Rev. Lett. \textbf{84}, 4019 (2000).

\bibitem{spin1chains}
C.~Hadley, A.~Serafini, and S.~Bose, Phys. Rev. A \textbf{72}, 052333 (2005). 

\bibitem{spin1chainsb}
D.~Burgarth and S.~Bose, Phys. Rev. A \textbf{71}, 052315 (2005). 

\bibitem{spin1chainsc}
O.~Romero-Isart, K.~Eckert, and A.~Sanpera, Phys. Rev. A \textbf{75}, 050303(R) (2007). 

\bibitem{spin1chainsd}
A.~Bayat and V.~Karimipour, Phys. Rev. A \textbf{75}, 022321 (2007).

\bibitem{spin1chainse}
K.~Eckert, O.~Romero-Isart, and A.~Sanpera, New J. Phys. \textbf{9}, 155 (2007).

\bibitem{bessen} A.~J.~Bessen, quant-ph/0609128.

\bibitem{abram} M.~Abramowitz and I.~A.~Stegun. \emph{Handbook of Mathematical Functions}. Dover, 1972.

\bibitem{braginsky} V.~B.~Braginsky, F.~Y.~Khalili and K.~S.~Thorne, Quantum Measurement, \emph{Cambridge University Press}, 1995.



\bibitem{quantumdots} R.~Hanson \emph{et al}. Rev. Mod. Phys. \textbf{79}, 1217 (2007).
\bibitem{cirac_porras} J.I.~Cirac and D.~Porras, Phys. Rev. Lett. \textbf{92}, 207901 (2004)
\end{thebibliography}
\end{document}